\documentclass[amsmath,superscriptaddress,citeautoscript,twocolumn,showpacs,floatfix]{revtex4}
\usepackage{amssymb}
\usepackage[pdftex]{graphicx}
\usepackage{mathrsfs}
\usepackage{amssymb,amsmath}
\usepackage{bm}
\usepackage{epsfig}
\usepackage{color}
\usepackage{amsfonts}
\usepackage{amscd}
\usepackage{leftidx}

\begin{document}

\title{Breakdown of Bose-Einstein distribution in photonic crystals}

\author{Ping-Yuan Lo}
\affiliation{Department of Physics and Center for Quantum Information 
Science, National Cheng-Kung University, Tainan 70101, Taiwan}

\author{Heng-Na Xiong}
\affiliation{Department of Physics and Center for Quantum Information 
Science, National Cheng-Kung University, Tainan 70101, Taiwan}

\author{Wei-Min Zhang}
\email{wzhang@mail.ncku.edu.tw}
\affiliation{Department of Physics and Center for Quantum Information 
Science, National Cheng-Kung University, Tainan 70101, Taiwan} 
\affiliation{National Center for Theoretical Science, Tainan 70101, Taiwan}

\begin{abstract}
In the last two decades, considerable advances  
have been made in the investigation of nano-photonics in
photonic crystals. Previous theoretical
investigations of photon dynamics were carried out at 
zero temperature. Here, we investigate micro/nano
cavity photonics in photonic crystals at finite temperature.  
Due to photonic-band-gap-induced non-Markovian 
dynamics, we discover that cavity 
photons in photonic crystals do not obey the standard 
Bose-Einstein statistical distribution. Within the photonic band 
gap and in the vicinity of the band edge, 
cavity photons combine nontrivial quantum dissipation with thermal 
fluctuations to form photon states that can memorize the initial cavity state 
information. As a result, Bose-Einstein distribution 
is completely broken down in these regimes, even if the thermal energy is  
larger than the photonic band gap. 
\end{abstract}
\pacs{ 03.65.Yz, 42.70.Qs, 05.70.Ln}

\maketitle

Photonic band gap (PBG) structures in photonic
crystals together with the characteristic dispersion properties make
light manipulations more efficient \cite{PC,manilightPC}. In
particular, light can be localized in photonic crystals due to PBG
\cite{localizedlight1,localizedlight2}, and ultrahigh quality-factor
cavities have been realized on-chip \cite{nanocavity-on-chip}.
Quantum optics with a few-level atom placed inside photonic crystals
has been extensively explored \cite{structured reservoir}. The
features of atomic population trapping (i.e. inhibition of spontaneous
emission) and atom-photon bound states in the vicinity of the
photonic band gap have been discovered \cite{QED-PBG1, John94,Kofman93}. 
These features are obtained mainly at zero temperature, solving 
the Schr\"{o}dinger equation in which the photonic crystal contains only
one single photon emitted from an atom which is initially in the
excited state. On the other hand, when the number of photons increases,
light propagating in photonic crystals was understood using the classical
Maxwell equations which are also defined at zero temperature \cite{PC,manilightPC}. 
Indeed, photonic quantum dynamics, even for a pure
micro/nano cavity embedded in photonic crystals, has not yet been 
solved at finite temperature. Practically,
understanding photonic quantum dynamics at finite temperature
is important for the development of all-optical circuits incorporating 
cavities and photonic bandgap waveguides in photonic crystals.

All-optical circuits for networks on-chips consist of micro/nano cavities 
and waveguides. Micro/nano cavities in photonic crystals are 
created by point defects, and photonic bandgap waveguides are made with coupled 
defect arrays.
Frequencies of the cavities and waveguides can be easily tuned by changing 
the size and/or the shape of defects.
To investigate the photon dynamics of micro/nano cavities which are coupled to
waveguides embedded in the photonic crystal at finite temperature,
one may treat both the photonic crystal and waveguides as reservoirs of  cavities.
Based on the recent development of the exact master equation for open
quantum systems \cite{PRL2012, Tu08235311, Jin09101765, Lei10}, 
an arbitrary photon state of  cavities connected 
through waveguides in photonic crystals
is governed by the following exact master equation \cite{Lei10}
\begin{align}
\dot{\rho} \left( t \right) =& - i  \left[H'_c(t), \rho \left( t \right) \right]  +\sum_{ij} 
\Big\{\kappa_{ij} \left( t \right) \big[ 2 a_j \rho \left( t \right) a_i
^{\dagger} \notag \\ & - a_i^{\dagger} a_j \rho \left( t \right) - \rho \left( t
\right) a_i^{\dagger} a_j \big] + \widetilde{\kappa}_{ij} \left( t \right) 
\big[ a_i^{\dagger} \rho \left( t \right) a_j \notag \\
& + a_j \rho \left( t \right) a_i^{\dagger} - a_i
^{\dagger} a_j \rho \left( t \right) - \rho \left( t \right) a_j a_i
^{\dagger} \big] \Big\}.
\label{exact_ME}
\end{align}
Here $\rho(t)$ is the reduced density matrix for cavity states, $a_i$ 
($a_i^\dag$) the photon annihilation (creation) operator,  $H'_c(t)=
\sum_{ij}\omega'_{c\,ij}(t) a^\dag_i a_j$ is the renormalized Hamiltonian of cavities 
with the environment-modifed cavity frequencies  $\omega'_{c\, ii}(t)$
and the environment-induced couplings between different cavities $\omega'_{c\, ij}(t)$,
after the environmental degrees of freedom are completely integrated out.
The coefficients $\kappa_{ij}
\left( t \right)$ and $\widetilde{\kappa}_{ij} \left( t \right)$ characterize
photon dissipations and fluctuations in photonic crystals at finite temperature.

\begin{figure}
\includegraphics[width=8.5cm]{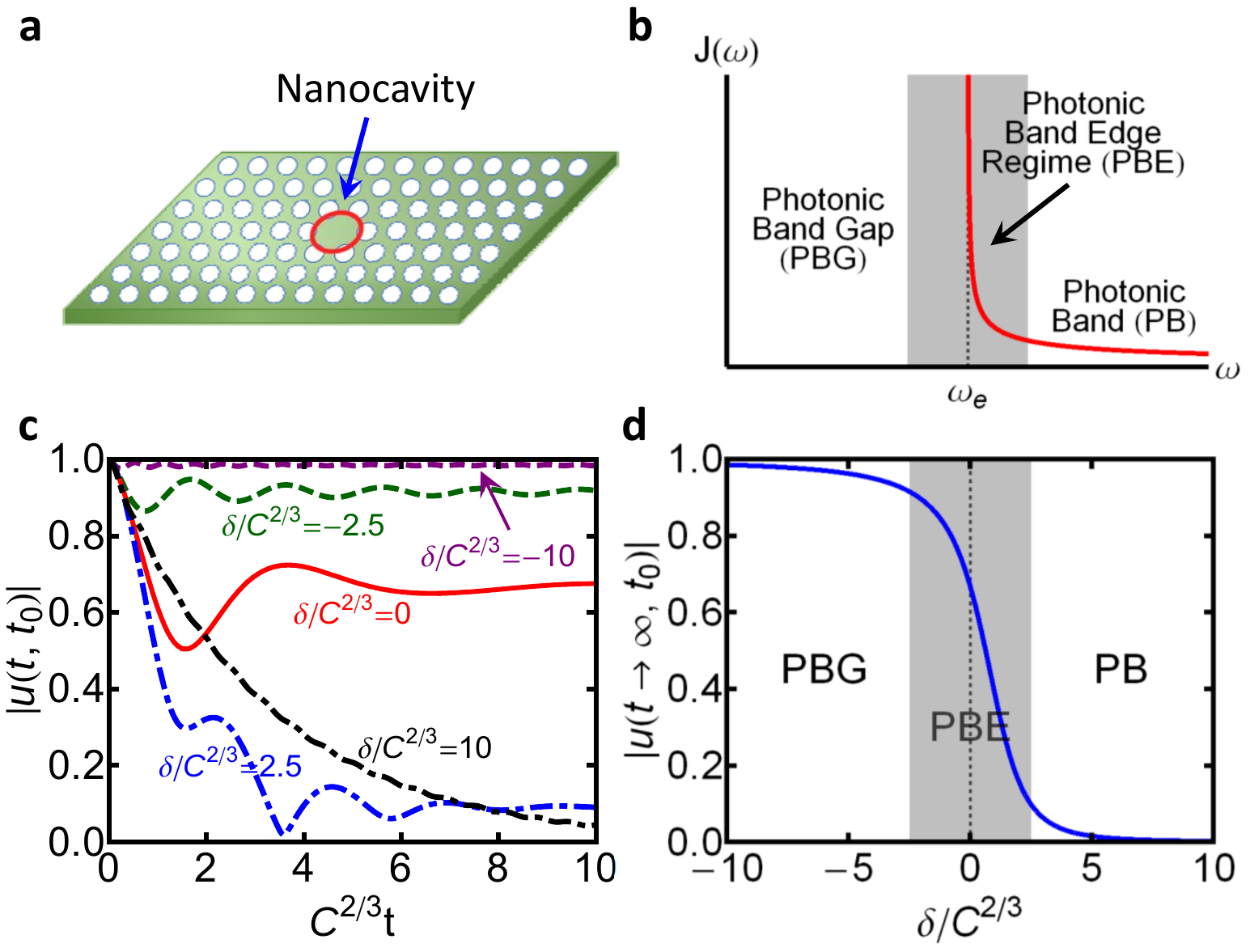}
\caption{
\textbf{a.} A micro/nano cavity embedded in photonic crystals. 
\textbf{b.} The spectral density $J \left( \omega \right)$ of the 
micro/nano cavity coupled to photonic crystals. $\omega = 
\omega_e$ is the photonic band edge. The spectrum is divided 
into three regimes: photonic band gap (PBG), the vicinity of the 
photonic band edge (PBE) and the photonic band (PB).   
\textbf{c.} Dissipative dynamics of the cavity field with different 
detuning $\delta= \omega _{c} - \omega _{e}$, with $\omega_e
=100 C^{2/3}$. \textbf{d.} Steady-state values of the normalized 
cavity field amplitude.
\label{fig1}}
\end{figure}

The master equation (\ref{exact_ME}) gives the exact
time-evolution of cavity photon states with arbitrary number of
photons in thermal photonic crystals.   To be specific, we may 
consider a single-mode micro/nano cavity 
in a photonic crystal, see Fig.~\ref{fig1}\textbf{a}.  The photon 
dissipations and fluctuations, given by the coefficients $\kappa \left( t \right)$
and $\widetilde{\kappa} \left( t \right)$ (all 
the sub-indices ($i,j$) in (\ref{exact_ME}) can be dropped now), are 
determined non-perturbatively and exactly by
nonequilibrium Green functions
\cite{Keldysh, Kad_Baym} through the relations
\cite{PRL2012, Tu08235311, Jin09101765, Lei10}:
$\kappa \left( t \right)+i\omega _{c} ^{\prime} \left( t \right) = -
\dot{u} \left( t, t_0 \right) u ^{-1} \left( t,t_0 \right)$, 
$\widetilde{\kappa} \left( t \right) = \dot{v} \left( t,t \right) + 2
v \left( t,t \right) \kappa \left( t \right) $. 
Here $u \left( t , t_0\right)$ is the photon field propagating Green
function: $\langle a(t) \rangle=u(t, t_0)\langle a(t_0) \rangle$, describing
cavity field relaxation, and $v \left( t ,t \right)$  
characterizes reservoir-induced photon thermal fluctuations 
and satisfies the relation with the average cavity photon number 
(cavity intensity):  $n \left( t \right) = \left\langle a ^{\dagger} \left( t
\right) a \left( t \right) \right\rangle = \left| u \left( t , t_0\right)
\right| ^{2} n \left( t _{0} \right) + v \left( t,t \right)$, where $\langle a(t_0) \rangle$
and $n(t_0)$ are the corresponding initial cavity field and initial 
cavity photon number.  These two Green functions are determined exactly
by the following Dyson equation and the nonequilibrium
fluctuation-dissipation theorem, respectively,
\begin{subequations}
\label{uv_eqns}
\begin{align}
& \dot{u}(t,t_0) \!= \!- i \omega _{c} u (t,t_0)\! - \!\! \int _{t _{0}} ^{t} \!\!\!dt g (t -t')
u (\tau,t_0) , \label{u_eqn} \\
& v (t,t) = \!\int^t_{t_0} \!\!\!d t_1\!\! \int _{t _0} ^{t}\!\!\!  d t_2 u^*(t_1,t_0) \widetilde{g}(t_1
-t_2) u(t_2,t_0) , \label{v_eqn}
\end{align}
\end{subequations}
where $\omega_c$ is the original cavity frequency.
The integral kernels in (\ref{uv_eqns}) characterize all the
back-actions between the cavity and its reservoirs, and
 are determined uniquely by the spectral density $J \left( \omega
\right)$ of reservoirs:
$g( t-t') \!\!= \!\!\int d \omega J(\omega ) e ^{- i \omega (t-t')}$,
and $\widetilde{g}(t-t')  \!\!= \!\! \int d \omega
J ( \omega ) \overline{n} ( \omega , T) e ^{-i \omega (t-t')}$,
where $\overline{n} \left( \omega , T \right) = \frac{1}{e ^{\hbar
\omega / k _{B} T} - 1}$ is the initial photon distribution in the
photonic crystal at temperature $T$.  The spectral density $J \left( \omega
\right)$ is microscopically defined as a multiplication of the density 
of states of the photonic crystal with the coupling 
strength between the cavity and the photonic crystal. In standard quantum optics, 
the spectral density is treated as a constant (white noise) so that the dissipation and fluctuation
coefficients, $\kappa$ and $\widetilde{\kappa}$, becomes time-independent 
\cite{Scully97}. The cavity state will then ultimately evolve into thermal 
equilibrium with its environment, and photons inside the cavity obey the
Bose-Einstein distribution \cite{Xiong13}. However, this well-known result can be
false for micro/nano cavities in thermal photonic crystals, as we will show below.

To investigate cavity photon state evolutions in photonic crystals, we must first solve 
the dynamics of photon dissipations and fluctuations, as determined by (\ref{uv_eqns}). 
In principle, the density of states of photonic crystals can be calculated  
numerically by solving photon eigenfrequencies and eigenfunctions 
in the photonic crystal with Maxwell's equations though 
the FDTD (finite-difference time-domain) method \cite{EM_theory}.  
For simplicity,  we consider an isotropic photonic crystal, and its density of state 
can be modeled \cite{QED-PBG1, John94} by $\rho _{PC} ( \omega ) \propto 
\frac{1}{\sqrt{\omega - \omega _{e}}} \Theta ( \omega - \omega _{e} )$,
where $\Theta ( \omega )$ is the Heaviside step function
and $\omega _{e}$ is the band edge, see Fig.~\ref{fig1}\textbf{b}.
Then the spectral density is given by 
\begin{equation}
J \left( \omega \right)= \frac{C}{\pi} \frac{1}{\sqrt{\omega - \omega
_{e}}} \Theta \left( \omega - \omega _{e} \right) , \label{spectrum}
\end{equation}
which could encapsulate the complete information about the effects of the reservoir
on photon dissipations and fluctuations \cite{Leggett87}. The parameter 
$C$ is a constant coupling strength between the cavity and photonic crystals, and thus
the cavity mode is assumed to couple equally with all possible modes in photonic crystals. 

With the spectral density (\ref{spectrum}), the cavity field propagating 
 function of (\ref{u_eqn}) can be solved exactly 
\begin{eqnarray}
u \left( t,t_0 \right) &=& \frac{2 \left( \omega _{e} - \omega _{b}
\right)}{3 \left( \omega _{e} - \omega _{b} \right) + \delta} 
e ^{- i \omega _{b} \left( t - t
_{0} \right)} \notag \\
&& + \frac{C}{\pi} \int _{\omega _{e}} ^{\infty} d \omega
\frac{\sqrt{\omega - \omega _{e}} e ^{- i \omega \left( t - t _{0}
\right)}}{\left( \omega - \omega _{c} \right) ^{2} \left( \omega -
\omega _{e} \right) + C ^{2}} , \label{u}
\end{eqnarray}
where $\delta = \omega _{c} - \omega _{e}$ is the detuning of
the cavity frequency from the photonic band edge (PBE). 
Equation (\ref{u}) shows that the photonic dissipation 
dynamics in photonic crystals contains two parts,  a localized photonic
mode (the first term) plus non-exponential damping (the second term).
The localized mode is dissipationless, induced by the PBG structure
in the photonic crystal.
The corresponding frequency $\omega _{b}$, which is the real root of the 
equation $(\omega _{c} - \omega _{b}) \sqrt{\omega _{e} - \omega _{b}}
=C$, lies within the PBG. The non-exponential damping comes from the
non-analyticity of the energy correction induced from the 
spectrum distribution profile of the photonic crystal. These results 
strongly rely on the detuning $\delta$, as shown in Fig.~\ref{fig1}\textbf{c}.

In fact, the localized mode and  the non-exponential damping 
indicates the existence of the non-Markovian memory dynamics \cite{PRL2012}.
As one can see in Fig.~\ref{fig1}\textbf{c}, when the
cavity mode is tuned far away from the PBG
($\delta \gtrsim 2.5 C ^{2 / 3}$), the optical field in the
cavity is rapidly damped, as shown in usual quantum optics. 
In the regime $-2.5 C ^{2 / 3} \lesssim \delta \lesssim 2.5 C ^{2 / 3}$, 
i.e. when the cavity frequency lies in the vicinity of the PBE, the 
non-exponential  non-Markovian damping dominates the
photon dynamics.  When
the cavity frequency is tuned deeply inside the PBG
($\delta \lesssim - 2.5 C ^{2 / 3}$), the cavity field has almost no 
damping, and thus light can be confined
in the defect of the photonic crystal, providing
a high-Q micro/nano cavity. In fact, these photon dissipative 
non-Markovian dynamics produces the same results with regard to atomic 
population trapping (inhibition of spontaneous emissions) and atom-photon
bound states in the vicinity of the photonic band gap, obtained by John
and others \cite{QED-PBG1, John94, Kofman93} when an atom 
is placed in the defect.  Figure \ref{fig1}\textbf{d} shows further 
the cavity field amplitude in the steady-state limit, due to the localized 
photon mode in (\ref{u}), 
as a dissipationless part (for $\delta \lesssim - 2.5 C ^{2 / 3}$). 
The localized photon mode decreases rather quickly 
near the PBE ($-2.5 C ^{2 / 3} \lesssim \delta \lesssim 2.5 C ^{2 / 3}$),
and becomes negligible for $\delta \gtrsim 2.5 C ^{2 / 3}$.  
To physically measure such non-Markovian dissipative dynamics in
photonic crystals, one may rewrite the solution (\ref{u}) in terms of 
the cavity field spectrum: $u (t,t_0) = \int d \omega \mathscr{D}
( \omega )e ^{- i \omega ( t - t _{0} )}$, with the reservoir-modified 
cavity field spectrum
\begin{align}
\mathscr{D} \left( \omega \right) =\mathscr{D} _{l}
 \left( \omega \right) + \mathscr{D} _{d} \left(
\omega \right) . \label{cDOS}
\end{align}
Here $\mathscr{D} _{l} ( \omega) = \mathcal{Z} \delta \left( \omega - \omega _{b} \right) =\frac{2 ( \omega _{e} - \omega
_{b})}{3( \omega _{e} - \omega _{b} ) + \delta}\delta ( \omega -
\omega _{b})$ and $\mathscr{D}_{d} ( \omega )= \frac{C}{\pi}
\frac{\sqrt{\omega - \omega _{e}}}{( \omega - \omega _{c}) ^{2} (
\omega- \omega _{e} ) + C ^{2}} $  denote the localized mode
and the continuous (dissipation) spectrum, respectively. 

With the above exact solution of photon dissipation dynamics, thermal 
photon fluctuations can be fully determined by 
the relation (\ref{v_eqn}), as a result of the generalized non-equilibirum 
fluctuation-dissipation theorem \cite{PRL2012}.
In solution (\ref{v_eqn}), the time correlation $\widetilde{g}(t-t')$
depicts the fluctuation due to the thermal photonic crystal.
When the system reaches its steady state, thermal fluctuations
are simply determined by the modified steady-state fluctuation-dissipation
theorem:
 $v ( t, t \rightarrow \infty ) = \int_{\omega_{e}}
^{\infty}  \mathcal{V} ( \omega ) d \omega$ with
\begin{align}
\mathcal{V} ( \omega ) = 
\overline{n} ( \omega , T ) \Big[ J ( \omega) \Big( \frac{\mathcal{Z}}
{\omega - \omega _{b}} \Big) ^{2} +\mathscr{D} _{d}
( \omega ) \Big] . \label{non_eq_FDT}
\end{align}
The first term is the localized mode contribution 
that modifies the conventional equilibrium fluctuation-dissipation
theorem. This additional contribution is
negligible when the cavity frequency is tuned far away from
the PBG, where $\omega_b \rightarrow \omega_e$ so that $\mathcal{Z} \rightarrow 0$.
Consequently, the solution (\ref{non_eq_FDT}) is reduced to the standard equilibrium
fluctuation-dissipation theorem: $\mathcal{V}( \omega ) =
\overline{n} ( \omega , T) \mathscr{D} _{d} ( \omega ) $.

\begin{figure} 
\includegraphics[width=8.5cm]{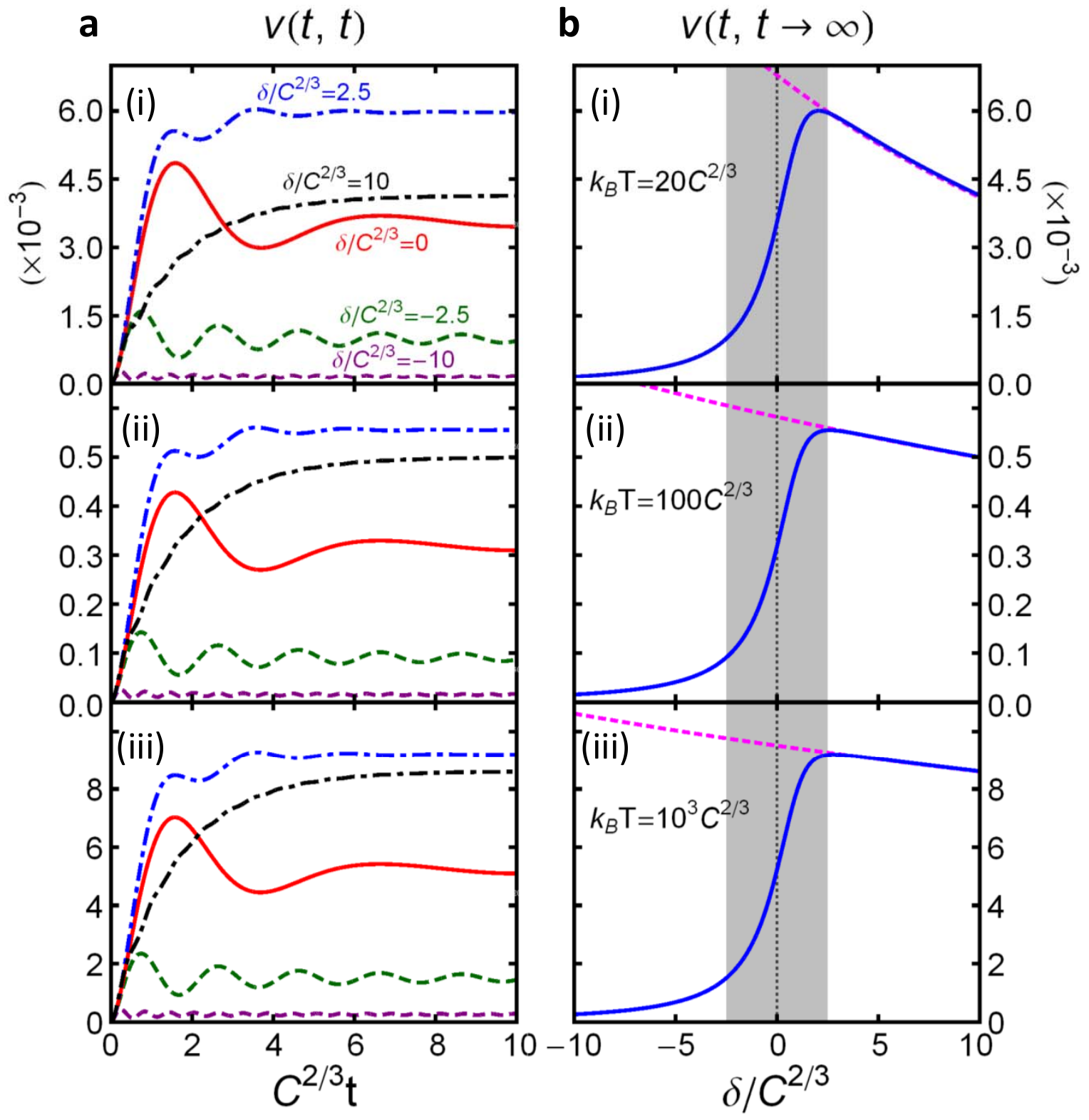}
\caption{
\textbf{a.} Time-evolution of thermal fluctuations, in terms of the 
correlation Green's function $v \left( t,t \right)$, due to the thermal 
photonic crystals. Different curves correspond to different detuning 
$\delta$, the same as that shown in Fig.~\ref{fig1}\textbf{c}, with 
the photonic crystal temperature (i) $k _{B} T = 20 C^{2/3}$, 
(ii) $k _{B} T = 100 C^{2/3}$ and (iii) $k _{B} T = 10^3 C^{2/3}$ 
and $\omega_e=100 C^{2/3}$. \textbf{b.} The steady-state values 
of thermal fluctuations. The solid-blue curve corresponds to the 
steady-state thermal fluctuation as a function of the detuning. 
The pink-dashed curve is the corresponding Bose-Einstein distribution. \label{fig2}}
\end{figure}

Thermal fluctuations at different initial temperatures
are presented in Fig.~\ref{fig2}\textbf{a}.
It appears that these fluctuations evolve in a similar way
for different temperatures. In the regime
where $\omega_c$ is tuned far away from PBG
($\delta \gtrsim 2.5 C ^{2 / 3}$), photons continuously flows
into the cavity from the photonic crystal until it reaches the
steady state. In this case, the cavity reaches thermal equilibrium
with the photonic crystal. When $\omega_c$ lies in
the vicinity of the PBE ($-2.5 C ^{2 / 3} \lesssim \delta
\lesssim 2.5 C ^{2 / 3}$), photons also flow into the cavity at
the beginning, but some of them are then transmitted back into the
photonic crystal, due to the non-Markovian memory effects. When
$\omega _{c}$ lies deeply inside the PBG ($\delta
\lesssim -2.5 C ^{2 / 3}$), few of the photons in the photonic crystal 
flow into the cavity, although the thermal energy, $k_BT =(20, 100, 10^3) C^{2/3}$
for the three graphs in Fig.~\ref{fig2}\textbf{a}, are larger or even much larger 
than the detuning $\delta$. Thermal fluctuations are thus suppressed significantly
in the PBG, due to the strong non-Markovian effect.

The physical picture of  thermal fluctuations can be seen 
by connecting it with the measurable average cavity photon number.
The average cavity photon number
is given by  $n \left( t \right)  = \left| u \left( t,t_0 \right)
\right| ^{2} n \left( t _{0} \right) + v \left( t,t \right)$, obtained directly from the master equation. 
This relation tells that $v(t,t)$ is just the thermal-fluctuation-induced 
average photon number in the cavity.  In Fig.~\ref{fig2}\textbf{b},
we present the steady-state thermal fluctuations.
The solid-blue curve is the thermal photon distribution of the photonic crystal at
given temperature. It shows that when $\delta \gtrsim 2.5 C ^{2 /
3}$, the thermal-fluctuation-induced steady-state average
photon number is identical to the thermal photon distribution in the photonic crystal:
$n(t \rightarrow \infty)= v(t, t \rightarrow \infty)=\bar{n}(\omega_c, T)
= 1/[e^{\hbar \omega_c/k_B T} -1]$, given by the dashed-pink curve.  
In this regime, the initial cavity photons are totally lost into the
photonic crystal as $u(t,t_0) \rightarrow 0$ in the steady-state limit,
as a typical Markovian process. 
The steady-state cavity photons all come from thermal fluctuations of the
reservoir.  As a consequence,  the cavity photon distribution obeys 
the Bose-Einstein distribution as that in the thermal photonic crystals. 
However, near the PBE ($-2.5C ^{2 /3} 
\lesssim \delta \lesssim 2.5 C ^{2 /3}$), 
the thermal-fluctuation-induced photon number 
deviates significantly from the thermal photon distribution, as shown
in Fig.~\ref{fig2}\textbf{b}.  In other words, Bose-Einstein distribution 
no longer works for cavity photons in the vicinity of 
the PBE.  When the cavity frequency $\omega_c$ lies deeply in the 
PBG ($\delta \lesssim -2.5 C ^{2 /3}$), the
thermal fluctuation effects approach to zero, and thus the cavity appears
isolated from the thermal reservoir and cannot reach equilibrium with the
reservoir \cite{Xiong13}, even if the thermal energy is larger or much larger 
than the detuning: $k_BT >\delta $ or $ k_BT >> \delta$. 
Bose-Einstein distribution is completely broken down here.

Based on the  above exact  photonic dissipation and fluctuation dynamics
in photonic crystals, we shall now solve the exact master equation and investigate
the photon state evolution. To be more specific,
consider the cavity to initially be  in a Fock state with an
arbitrary photon number $n_0$, i.e.
$\rho \left( t _{0} \right) = \left| n _{0} \right\rangle
\left\langle n _{0} \right|$, which can be prepared experimentally 
\cite{Haroche11}. By solving the master equation, the 
cavity state at arbitrary time $t$ is given by
$\rho (t)= \sum \limits _{n = 0} ^{\infty}
\mathcal{P} _{n} ^{( n _{0} )} (t) | n \rangle \langle n | $ with
\begin{align}
\mathcal{P} _{n} ^{( n _{0})} (t) =& \frac{[ v (t,t)] ^{n}}{[ 1 +
v (t,t) ] ^{n + 1}} \left[ 1 - \Omega \left( t \right) \right] ^{n _{0}}\notag \\
& \times \!\!\!\sum_{k=0}^{\rm min\{n_0,n\}}\!\!\!\Bigg(\!\!\begin{array}{c} n_0 \\ k \end{array}\!\!\Bigg)
\Bigg(\!\!\begin{array}{c} n \\ k \end{array}\!\!\Bigg)\Bigg[\frac{1}{v(t,t)}\frac{\Omega(t)}{1-\Omega(t)}
\Bigg]^k
 \label{Fock_P}
\end{align}
where $\Omega \left( t \right) = \frac{\left| u \left( t ,t_0\right) \right| ^{2}}{1 + v \left( t ,t\right)}$.
This result shows that an initial photon Fock state will evolve
into a mixed state of different Fock states, and the probability
in each Fock state $\left| n \right\rangle$ is $\mathcal{P}
_{n} ^{\left( n _{0} \right)} \left( t \right)$.

\begin{figure}
\includegraphics[width=8.5cm]{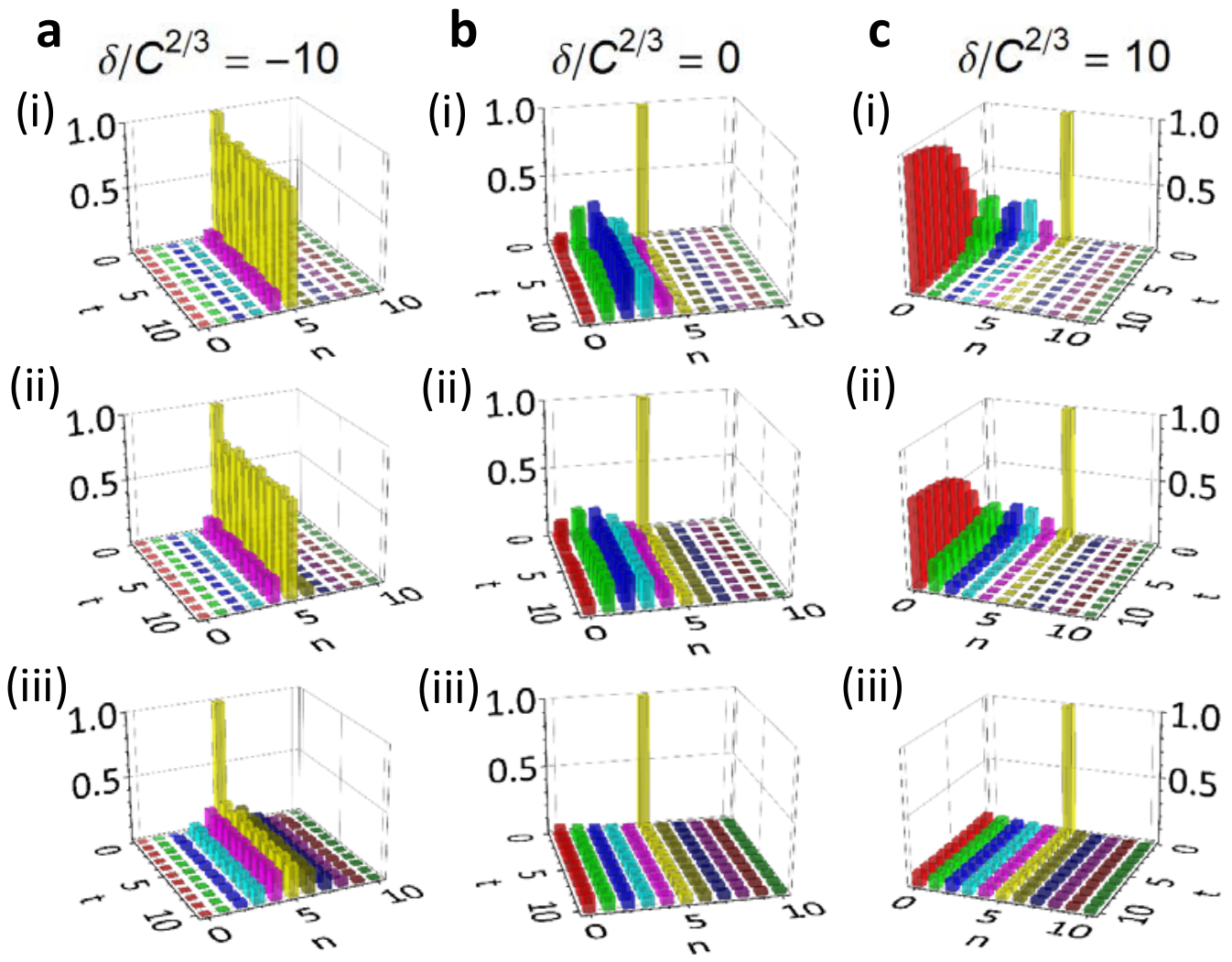}
\caption{
The time-evolution of the probability distribution,
$\mathcal{P} _{n} ^{\left( n _{0}\right)} \left( t \right)$ for different detuning $\delta$ and 
different temperature $T$: (i) $k _{B} T = 20 C^{2/3}$
(ii) $k _{B} T = 100 C^{2/3}$, and (iii) $k _{B} T = 10^3 C^{2/3}$, 
with the cavity being  in an initial Fock state $|n_0\rangle$ 
with $n_0=5$ and the band edge $\omega_e=100 C^{2/3}$. 
The time $t$ is in the unit of $C^{-2/3}$. \label{fig3} }
\end{figure}

The time-evolution of the probability distribution
$\mathcal{P} _{n} ^{( n _{0})} (t)$ for the initial 
state $\left| n _{0} = 5 \right\rangle$ is given in Fig.~\ref{fig3}.
The results show that inside the PBG, e.g.
$\delta = -10 C ^{2 / 3}$, and when the photonic
crystal temperature is not too high ($k _{B} T = 20 C ^{2 / 3}$ 
and  $100 C ^{2 / 3}$), the cavity maintains  the initial Fock state. It only has a small chance
to decay to the Fock state $\left| n = 4 \right\rangle$ [see Fig.~\ref{fig3}\textbf{a}-(i)
with $k _{B} T = 20 C ^{2 / 3}$], and an even smaller
probability to be in the Fock state $\left| n = 6 \right\rangle$
[see Fig.~\ref{fig3}\textbf{a}-(ii) when $k _{B} T$ increases to 
$100 C ^{2 / 3}$]. However, at a high temperature ($k _{B} T = 
10^3 C ^{2 / 3}$) the thermal fluctuation becomes relatively strong, such that most of
the initial state information will be lost and the cavity
evolves to a mixed state covering several Fock states around the
initial one $\left| n _{0} = 5 \right\rangle$ [see
Fig.~\ref{fig3}\textbf{a}-(iii)]. Near the PBE (e.g. $\delta = 0$), the cavity will lose
photons into the photonic crystal relatively easily. At a low temperature
($k _{B} T = 20C ^{2 / 3}$), the photon loss makes the
cavity become a mixed state of several Fock states $\left| n
\right\rangle$ only for $n < 5$, and mainly distributed around $n = 2$
and $3$, see Fig.~\ref{fig3}\textbf{b}-(i). At a relatively high temperature ($k
_{B} T =100C ^{2 / 3}$), the cavity state also becomes a mixed state
of several Fock states $\left| n \right\rangle$, distributed mainly
among $n = 1$ to $3$, but the distribution is broader, 
see Fig.~\ref{fig3}\textbf{b}-(ii). As the temperature
becomes rather high ($k _{B} T = 10^3 C ^{2 / 3}$), the
time-evolution of the cavity state behaves quite differently
because thermal fluctuations now play a crucial role, 
and thus the structure of the initial state is quickly destroyed, as shown in
Fig.~\ref{fig3}\textbf{b}-(iii). However, the cavity does not really reach
thermal equilibrium with the photonic crystal, 
as can be seen from Figs.~\ref{fig1}\textbf{d} and \ref{fig2}\textbf{b}, in which
$\Omega(t)=\frac{\left| u \left( t,t_0 \right) \right| ^{2}}{1 + v \left( t,t \right)} \lesssim 0.1$.
Taking the solution (\ref{Fock_P}) up to the order of $\Omega (t)$, we have
\begin{equation}
\mathcal{P} _{n} ^{\left( n _{0} \right)} \left( t \right) \approx
\frac{\left[ v \left( t,t \right) \right] ^{n}}{\left[ 1 + v \left( t,t
\right) \right] ^{n + 1}} \!\! \left[ 1 \!\! - \!\! \left( 1\!\! - \!\! \frac{n}{v \left( t,t
\right)} \right) \! \Omega \left( t \right) n _{0} \right] . \label{Fock_P_approx}
\end{equation}
Here the second term shows the deviation from the thermal-like
state $\rho _{T} \left[ v \left( t ,t\right) \right] = \sum \limits _{n = 0}^{\infty} 
\frac{\left[v \left( t,t \right) \right] ^{n}}{\left[ 1 + v \left( t,t \right) \right] ^{n + 1}} 
\left| n \right\rangle \left\langle n \right|$, where $v \left( t,t
\right)$ is given by Eq.~(\ref{v_eqn}) [its steady-state limit is given 
by Eq.~(\ref{non_eq_FDT})]. In this case ($\delta = 0$ and $k _{B} T =
10^3 C ^{2 / 3}$), the cavity evolves into a mixed state
slightly different from the thermal-like one $\rho _{T} \left[ v
\left( t ,t\right) \right]$. If we turn $\omega_c$ into the PB (e.g. $\delta = 10 C ^{2 / 3}$), 
then $u(t,t_0) $ will gradually decay to zero, indicating that the
cavity gradually emits all photons into the photonic crystal and
finally becomes a vacuum at very low temperature, see
Fig.~\ref{fig3}\textbf{c}-(i). At higher temperature, the thermal fluctuation 
$v(t,t)$ approaches to the Bose-Einstein distribution: $v(t,t \rightarrow
\infty)= \overline{n} \left( \omega _{c} , T \right)$. Only in this case,
the cavity truly evolves into thermal equilibrium with the photonic 
crystal, given by $\mathcal{P} _{n} ^{\left( n _{0} \right)} \left( t
\right) \rightarrow \frac{\left[ \overline{n} \left( \omega _{c} , T
\right) \right] ^{n}}{\left[ 1 + \overline{n} \left( \omega _{c} , T
\right) \right] ^{n + 1}}$, as shown in Fig.~\ref{fig3}\textbf{c}-(ii) and
-(iii).

\begin{figure}
\includegraphics[width=8.5cm]{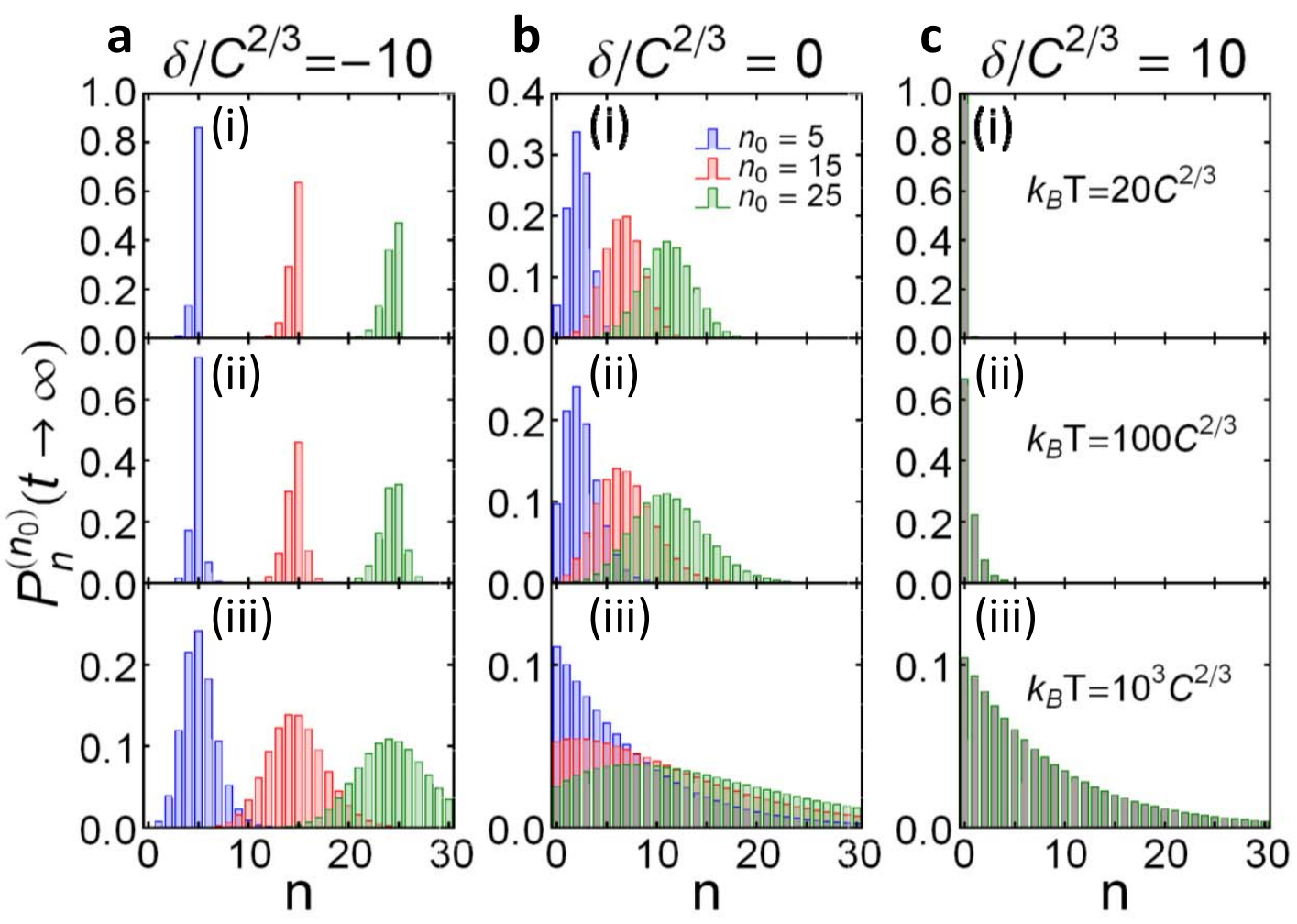}
\caption{
The steady-state cavity photon distribution, $P _{n} ^{\left( n _{0} 
\right)} \left( t\rightarrow \infty \right)$ for different initial states 
$|n_0\rangle$ with $n _{0} = 5, 15, 25$, and for different temperatures 
$T$ (i) $k _{B} T = 20C^{2/3}$, 
(ii) $k _{B} T = 100C^{2/3}$ and (iii) $k _{B} T = 10^3C^{2/3}$, 
and different detuning \textbf{a.} $\delta = -10C ^{2 / 3}$  (inside 
the PBG regime), where the Bose-Einsten distribution is completely 
broken down; \textbf{b.} $\delta = 0$ (at the band edge), where the 
cavity steady states still depend on the initial state and do not obey 
Bose-Einstein distribution, and \textbf{c.} $\delta = 10 C ^{2 / 3}$ 
(within the PB regime), in this case, the cavity steady states show 
the Bose-Einstein distribution for all different temperatures. \label{fig4} }
\end{figure}
The initial state dependence can be further
examined from the steady-state limit of the cavity photon distribution
$\mathcal{P} _{n} ^{\left( n _{0} \right)} \left( t \rightarrow
\infty \right)$. The results are presented in Fig.~\ref{fig4}.
In each graph in Fig.~\ref{fig4}, we take three different initial Fock States $\left| n
_{0} \right\rangle$. Figure~\ref{fig4} shows that inside the PBG
($\delta = - 10 C ^{2 / 3}$) the steady state cavity photon distribution 
 strongly depends on the initial state, as a significant evidence of non-Markovian dynamics. At low
temperature (e.g. $k _{B} T = 20 C ^{2 / 3}$), the cavity photons are distributed
in the regime $n \leq n _{0}$ with the maximum peak at $n = n _{0}$, see
the first graph in Fig.~\ref{fig4}\textbf{a}. At a higher temperature (e.g. 
$k _{B} T = 100C ^{2 / 3}$ and $10^3C ^{2 / 3}$), the cavity photons
are distributed slightly more broadly, due to thermal fluctuations, but are still
centered around the initial photon number $n _{0}$, as seen in
the second and the third graphs in Fig.~\ref{fig4}\textbf{a}.  This is because 
the non-Markovian memory still dominates the quantum dynamics.
In these cases, Bose-Einstein distribution completely breaks down.
Near the PBE ($\delta = 0$), the cavity photons are
centered at a number slightly less than $n _{0} / 2$ if the
temperature is not too high ($k _{B} T = 20 C ^{2 / 3}$ and
$100 C ^{2 / 3}$), as seen in the first two graphs in Fig.~\ref{fig4}\textbf{b}. 
At a high temperature ($k _{B} T = 10^3C ^{2 / 3}$),   thermal fluctuations 
play a more important role in the cavity dynamics
so that the steady cavity states for different initial
states do not differ from each other as that much as the previous cases, 
but the photon distribution still derivates significantly from the standard 
Bose-Einstein distribution due to the non-Markovian effect, as seen in the 
last graph in Fig.~\ref{fig4}\textbf{b}, 
as a result of the solution to (\ref{Fock_P_approx}).
When $\omega_c$ lies far away from the PBG (e.g. $\delta = 10 C ^{2 / 3}$), 
photons in the cavity will gradually be damped 
into the photonic crystal, and the photons in the photonic crystal are 
transferred into the cavity through thermal fluctuations.  In this case
the non-Markovian effect is negligible, and the cavity ultimately reaches thermal equilibrium with the 
photonic crystal, and the Bose-Einstein statistical distribution is recovered,
$\mathcal{P} _{n} ^{\left( n _{0} \right)} \left( t\rightarrow \infty
\right) = \frac{\left[ \overline{n} \left( \omega _{c} , T
\right) \right] ^{n}}{\left[ 1 + \overline{n} \left( \omega _{c} , T
\right) \right] ^{n + 1}}$, where the initial state 
information is completely washed out, as shown in Fig.~\ref{fig4}\textbf{c}. 

In conclusion, we show that when the cavity frequency lies
near the photonic band edge or deeply inside the photonic band gap in photonic crystals, 
the steady cavity states do not obey the Bose-Einstein statistical distribution, even though the 
thermal energy is larger or much larger than the detuning.  This non-trivial photon 
dynamics provides a direct evidence for the impact of non-Markovian dynamics 
on the equilibrium hypothesis of statistical mechanics \cite{Xiong13}, and it 
could be examined with current photonic crystal experiments  in microwave regime.

This work is supported by the National Science
Council of Republic of China under Contract No. NSC-102-2112-M-006-016-MY3.
\ \\

\noindent \textbf{References} \\


\ \\


\end{document}